\documentclass[%
reprint,
superscriptaddress,
showpacs,
preprintnumbers,
 bibnotes,
 amsmath,amssymb,
 aps,
prb,
]{revtex4-1}

\usepackage{graphicx}% Include figure files
\usepackage{dcolumn}% Align table columns on decimal point
\usepackage{bm}% bold math
\usepackage{color}
\begin{document}

%\preprint{Preprint}

\title{Growth and characterization of HgBa$_{2}$CaCu$_{2}$O$_{6+\delta}$ and HgBa$_{2}$Ca$_{2}$Cu$_{3}$O$_{8+\delta}$ crystals}

\author{Lichen~Wang}
\affiliation{International Center for Quantum Materials, School of Physics, Peking University, Beijing 100871, China}
\author{Xiangpeng~Luo}
\altaffiliation[Present address: ]{Department of Physics, University of Michigan, Ann Arbor, Michigan 48109, USA}
\affiliation{International Center for Quantum Materials, School of Physics, Peking University, Beijing 100871, China}
\author{Jiarui~Li}
\altaffiliation[Present address: ]{Department of Physics, Massachusetts Institute of Technology, Cambridge, Massachusetts 02139, USA}
\affiliation{International Center for Quantum Materials, School of Physics, Peking University, Beijing 100871, China}
\author{Junbang~Zeng}
\affiliation{International Center for Quantum Materials, School of Physics, Peking University, Beijing 100871, China}
\author{Minghao~Cheng}
\altaffiliation[Present address: ]{Department of Physics, Columbia University, New York 10027, USA}
\affiliation{International Center for Quantum Materials, School of Physics, Peking University, Beijing 100871, China}
\author{Jacob~Freyermuth}
\affiliation{School of Physics and Astronomy, University of Minnesota, Minneapolis, Minnesota 55455, USA}
\author{Yang~Tang}
\affiliation{School of Physics and Astronomy, University of Minnesota, Minneapolis, Minnesota 55455, USA}
\author{Biqiong~Yu}
\affiliation{School of Physics and Astronomy, University of Minnesota, Minneapolis, Minnesota 55455, USA}
\author{Guichuan~Yu}
\affiliation{School of Physics and Astronomy, University of Minnesota, Minneapolis, Minnesota 55455, USA}
\author{Martin~Greven}
\affiliation{School of Physics and Astronomy, University of Minnesota, Minneapolis, Minnesota 55455, USA}
\author{Yuan~Li}
\email[]{yuan.li@pku.edu.cn}
\affiliation{International Center for Quantum Materials, School of Physics, Peking University, Beijing 100871, China}
\affiliation{Collaborative Innovation Center of Quantum Matter, Beijing 100871, China}

\begin{abstract}
We report the successful synthesis of single-crystalline cuprate superconductors HgBa$_{2}$CaCu$_{2}$O$_{6+\delta}$ and HgBa$_{2}$Ca$_{2}$Cu$_{3}$O$_{8+\delta}$. These compounds are well-known for their high optimal superconducting critical temperatures of $T_\mathrm{c}$ = 128 K and 134 K at ambient pressure, respectively, and for their challenging synthesis. Using a conventional quartz-tube encapsulation method and a two-layer encapsulation method that utilizes custom-built high-pressure furnaces, we are able to grow single crystals with linear dimensions up to several millimeters parallel to the CuO$_2$ planes. Extended post-growth anneals are shown to lead to sharp superconducting transitions, indicative of high macroscopic homogeneity. X-ray diffraction and polarized Raman spectroscopy are identified as viable methods to resolve the seemingly unavoidable inter-growth of the two compounds. Our work helps to remove obstacles toward the study of these model cuprate systems with  experimental probes that require sizable high-quality crystals.

\end{abstract}

%\pacs{74.70.Xa, %Pnictides and chalcogenides
%78.70.Nx  %Neutron inelastic scattering
%}

\maketitle

\section{INTRODUCTION}

The high-$T_\mathrm{c}$ superconducting cuprates (HTSCs) have attracted a tremendous amount of interest for their significance in fundamental science and their potential for applications. Despite enormous research efforts since the original discovery of HTSCs more than three decades ago \cite{JGZPRB1986}, the microscopic mechanism that gives rise to superconductivity remains one of most important unresolved questions in condensed matter physics research. The HTSCs are doped Mott insulators and, as a result of strong quasi-two-dimensional electronic correlations, exhibit a rich variety of phases as a function of temperature, doping, pressure, and magnetic field \cite{KeimerNature2015,VojtaAdvinPhys2009,ArmitageRevModPhys2010,TranquadaPhyB2015}. Whereas each family of compounds possesses some specific properties, superconductivity is thought to originate from the pivotal structural and electronic unit shared by all HTSCs: the CuO$_2$ plane. Hole-doped compounds exhibit higher $T_c$ values than their electron-doped counterparts, and the highest transition temperatures correspond to a doping level of $p \sim 0.16$ holes per planar Cu (``optimal doping''). Many intriguing phases and ordering tendencies are found in the ``underdoped'' regime, as the system evolves from the undoped insulator toward optimal doping. The superconducting phase is stabilized between $p \sim 0.05$ and $\sim 0.25$, and highly ``overdoped'' compounds exhibit conventional metallic (Fermi-liquid) behavior with no apparent signatures of the Mott-insulating state.

\begin{figure}
\includegraphics[width=3.375in]{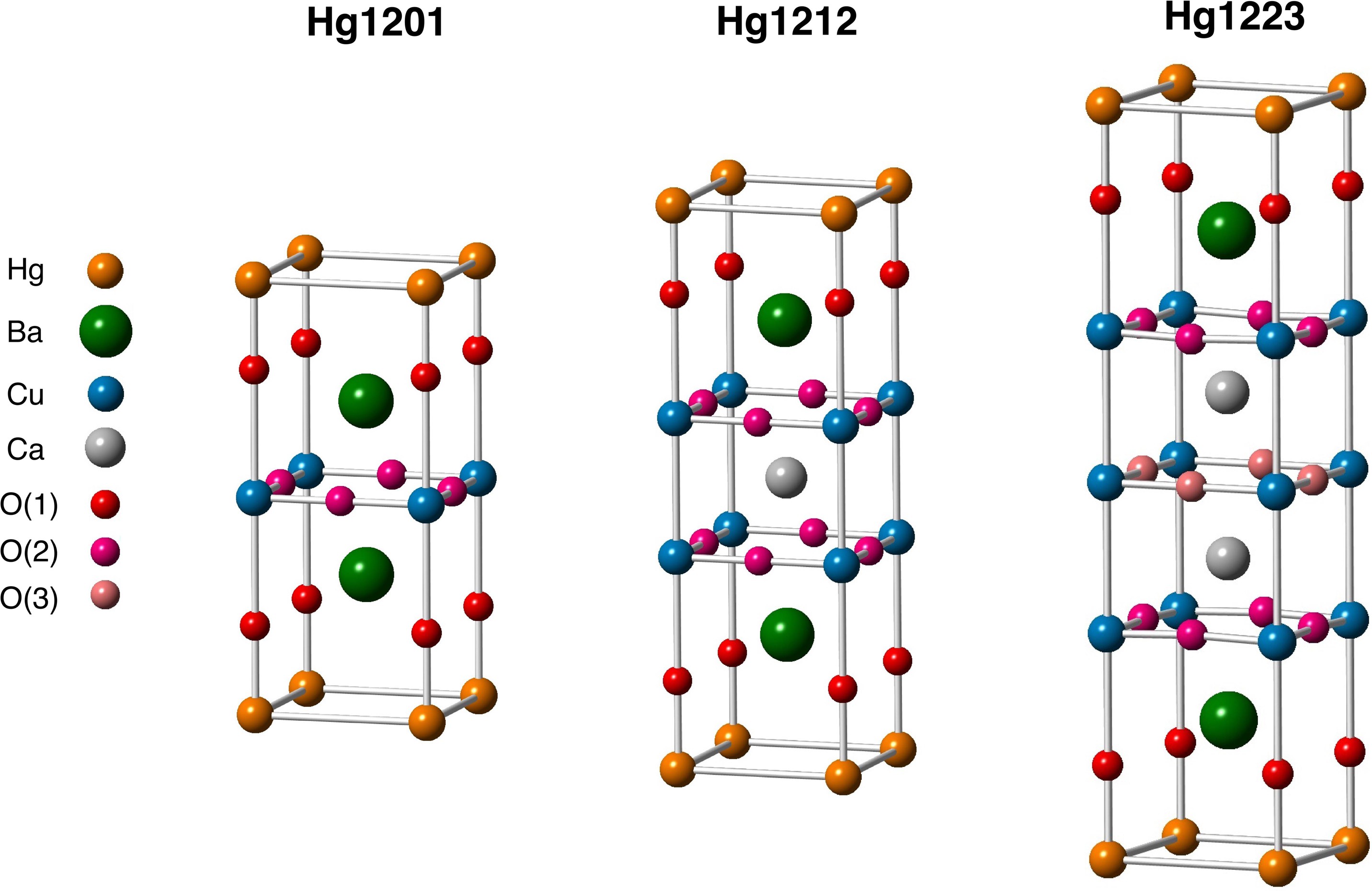}
\caption{\label{Figbbb}
Structural comparison of Hg-family cuprates.
}
\end{figure}

Among the several hundred cuprates, those compounds with high structural symmetry and higher $T_\mathrm{c}$ are commonly considered advantageous for experimental study. Low structural symmetry can mask the underlying behavior of the CuO$_2$ planes \cite{ChanPRL2014} and low $T_\mathrm{c}$ values are often associated with stronger point disorder effects and/or stronger competing phases \cite{EisakiPRB2004}. Consequently, the Hg-family of compounds, whose general chemical formula reads HgBa$_{2}$Ca$_{n-1}$Cu$_{n}$O$_{2n+2+\delta}$, has received considerable attention. Here, $n$ indicates the number of CuO$_2$ layers in the simple tetragonal primitive cell, and $\delta$ indicates the concentration of oxygens in the Hg-O layer. The structures of the first three members of this family are shown schematically in Fig.~\ref{Figbbb}.
The maximal value of $T_\mathrm{c}$, fully optimized as a function of $\delta$ (and hence $p$), depends on $n$: it increases from 97 K ($n=1$, Ref. \onlinecite{PutilinNature1993}), to 127 K ($n=2$, Ref. \onlinecite{SchillingNature1993,PutilinPhysC1993}), to 134 K ($n=3$, Ref. \onlinecite{SchillingNature1993}), and then decreases again at even larger $n$ (Ref. \onlinecite{AntiSupSciTec2002}).
The Hg-family cuprates exhibit the highest optimal $T_\mathrm{c}$ values among all HTSCs with the same number of CuO$_{2}$ layers in the primitive cell \cite{EisakiPRB2004}, and indeed the $T_\mathrm{c}$ value of the $n=3$ compound is the highest among all HTSCs at ambient pressure. In addition, the Hg-family cuprates exhibit high tetragonal symmetry (with no known structural phase transitions) and chemical disorder is confined to the Hg-O layers, which are relatively far away from the CuO$_{2}$ layers \cite{EisakiPRB2004}. In the case of single-layer HgBa$_2$CuO$_{4+\delta}$ (Hg1201), it has been demonstrated that a rather wide hole-doping range is accessible \cite{YamamotoPRB00,BarisicPRB2008} and that disorder effects on the electronic properties of the the CuO$_{2}$ layers are minimal \cite{BarisicPNAS2013,ChanPRL2014,BarisicNatPhys2013,ChanNatCommun2016}. Therefore, the Hg-family cuprates can be considered model systems and the most promising choice for experimental studies. Experiments on these materials had been rather limited due to the lack of high-quality single crystals until the breakthrough in the crystal growth of the single-layer Hg1201 that employed an encapsulation method \cite{ZhaoAdv2006}. Based on such crystals, many state-of-the-art measurements have been performed, including
magnetic neutron scattering \cite{LiNa2008,YuPRB2010,LiNature2010,Li2011,LiPRB2011,LiNatPhys2012,Chan2016a,Chan2016,Tang2018},
charge transport \cite{BarisicPRB2008, BarisicPNAS2013,ChanPRL2014,BarisicNatPhys2013,ChanNatCommun2016,Grbic2009,Doiron-Leyraud2013,Pelc2017,Popcevic2017},
thermodynamic \cite{Hardy2010,Yu2017,Murayama2018} and nuclear magnetic resonance \cite{Rybicki2009,Rybicki2015,Haase2012} experiments,
as well as the determination of charge correlations and excitation properties with a variety of photon-based techniques \cite{HomesNature2004,LuPRL2005,HeumenPRB2007,HeumenPRB2009,LiPRL2012,LiPRL2013,MirzaeiPNAS2013,VishikPRB2014,Wang2014,CilentoNatCommun2014,TabisNatCom2014,HintonSciRep2016,TabisPRB2017}.
Key results include the demonstration of universal translational-symmetry-preserving magnetism \cite{LiNa2008,Li2011} and quantum oscillations \cite{BarisicNatPhys2013,ChanNatCommun2016} in the pseudogap state below optimal doping, and the demonstration that the magnetoresistance obeys Kohler scaling, with a Fermi-liquid scattering rate \cite{ChanPRL2014}.

In the present study, we extend the growth and characterization of sizable single crystals to the multi-layered sister compounds HgBa$_{2}$CaCu$_{2}$O$_{6+\delta}$ (Hg1212) and HgBa$_{2}$Ca$_2$Cu$_3$O$_{8+\delta}$ (Hg1223). Part of the motivation is to enable the systematic study of the most desirable, highest-$T_\mathrm{c}$ cuprates. Comparison with the established properties of single-layer Hg1201 can be expected to be illuminating. Moreover, Hg1212 is of high interest for a direct comparison with YBa$_2$Cu$_3$O$_{6+x}$, arguably the most studied HTSC, which features CuO chains in addition to the CuO$_2$ bilayers and lower, orthorhombic symmetry in the superconducting doping range. Finally, Hg1223 offers an opportunity to investigate a cuprate system with of two inequivalent types of CuO$_{2}$ layers: the outer layers are adjacent to the ``charge reservoir'' Hg-O layers, whereas the inner layer is partially ``screened'' by the outer layers and may hence be less doped \cite{KotegaJPhy2001}. Altogether, we believe that research on Hg1212 and Hg1223 will facilitate a deeper understanding of existing results on hole-doped HTSCs and potentially enable new discoveries. Here we report the successful synthesis of sizable single crystals of Hg1212 and Hg1223, along with basic sample characterization results, including magnetic susceptibility, X-ray diffraction and Raman spectroscopy.

\section{SYNTHESIS}

The synthesis of Hg-family of HTSCs presents a major challenge, especially for the growth of large single crystals, because mercury is volatile. The existence of gaseous Hg contrasts with the syntheses of many other HTSCs that mainly involve solid-state or solid-liquid-state reactions. In order to ensure a sufficiently high density of gaseous Hg, high-pressure techniques have been applied since the discoveries of the Hg-family HTSCs. Extremely high pressures up to 60 Kbar (1 bar = 10$^{5}$ Pa), generated by a belt-type apparatus, were originally used to synthesize polycrystalline samples \cite{PutilinPhC1993,AntipovPhC1993}. Although such techniques were effective, especially for the synthesis of the multi-layered members of the Hg-family of HTSCs, these approaches precluded the growth of sizable single crystals due to the limitations on sample volume. A different approach that utilized excess Hg vapor pressure was also effective in suppressing the evaporation of Hg and in improving the phase stability of the desired product \cite{AntiSupSciTec2002, AlyoshinPhysC1996, MengPhysC1993, ColsonPhysC1994, XuePhysC1997, KnizekPhC2002}. The required Hg pressures were reported to range from a few bars to a few tens of bars \cite{AntiSupSciTec2002, XuePhysC1997, KnizekPhC2002}, which was much more compatible with a large sample volume, making it feasible to grow sizable single crystals.

Early syntheses were mostly performed at temperatures around 800 $^{\circ}$C (Refs.~\onlinecite{PutilinPhC1993, AntipovPhC1993, AlyoshinPhysC1996, MengPhysC1993, ColsonPhysC1994, XuePhysC1997, KnizekPhC2002}), at which the reactions were of a solid-gas-state nature, which usually resulted in an end product of powder form.  Although in some reports crystals up to sub-millimeter size in CuO$_2$-planar dimensions were obtained \cite{ColsonPhysC1994}, the growth of even larger crystals appeared to be out of reach. Researchers subsequently used flux methods, which required heating the reacting materials to above 1000 $^{\circ}$C (Refs.~\onlinecite{ZhaoAdv2006,PelloquinPhysC1997,WisniewskiPRB2000}). In particular, Zhao and coworkers \cite{ZhaoAdv2006} succeeded in obtaining Hg1201 single crystals with very large volumes up to $\sim 100$ mm$^{3}$, thanks to the fact that the phase stability of Hg1201 requirses the lowest Hg pressure among all members in the family\cite{AntiSupSciTec2002, XuePhysC1997, KnizekPhC2002}. The method involved the use of thick-walled quartz tubes to sustain  vapor pressures of about 10 bar at above 1000 $^{\circ}$C, a temperature at which the flux was in molten form, so that subsequent crystallization could take place from a relatively large amount of liquid upon slow cooling.

\begin{figure}
\includegraphics[width=3in]{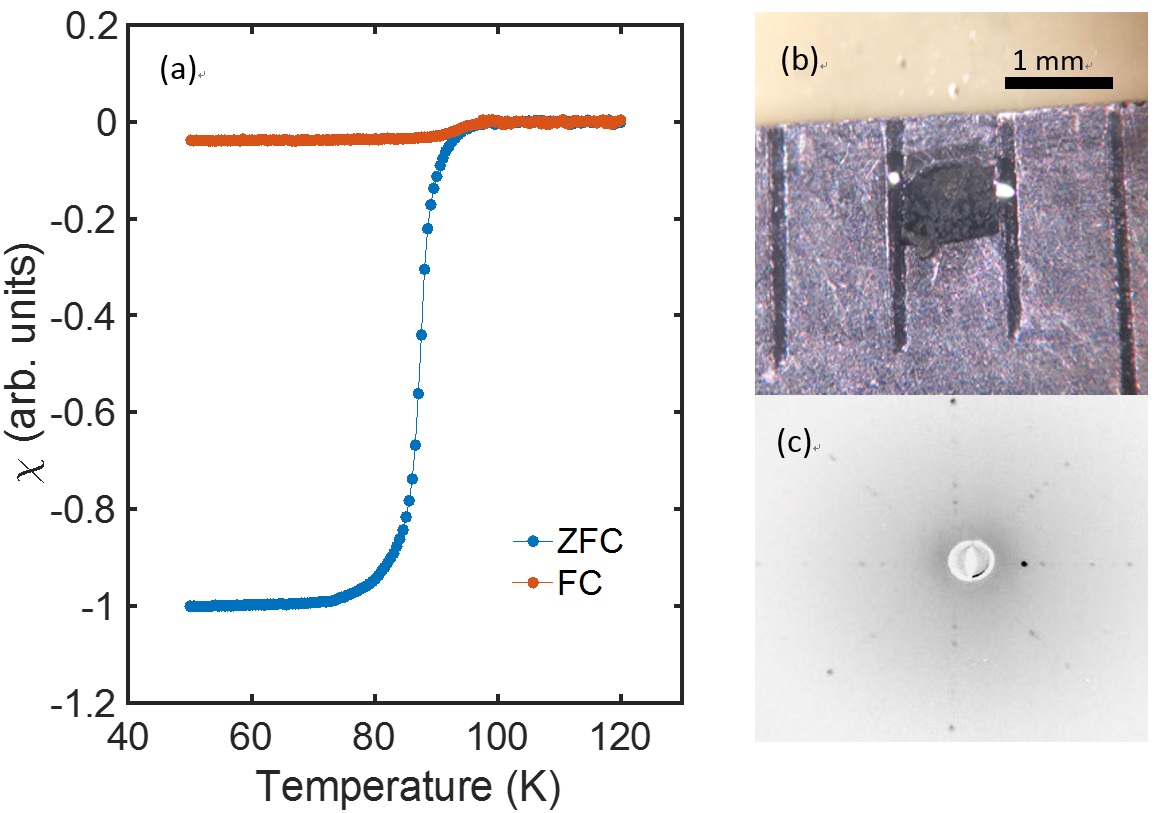}
\caption{\label{Fig0}
(a) Magnetic susceptibility measurement of a Hg1212 crystal grown with the conventional method (see text) in a 5 Oe $c$-axis field, upon cooling in a magnetic field (FC) and after cooling in zero field (ZFC). (b) Photograph of the crystal.  (c) $c$-axis X-ray Laue pattern obtained for the crystal.
}
\end{figure}

Our first attempt to grow single crystals of double-layer Hg1212 was performed by modifying the method of Zhao \textit{et al}.\cite{ZhaoAdv2006} Stoichiometric precursor starting materials containing Ba(NO$_3$)$_2$ and CuO were sintered at 920 $^{\circ}$C in oxygen flow, followed by mixing with excess amounts (30\% more than the stoichiometric amount) of CaO powder using a mortar and pestle. This mixture was then placed inside a zirconia crucible, along with an excess amount of HgO powder (60\% more than the stoichiometric amount), and sealed inside an evacuated quartz tube with a small amount (20-30 mg) of MgSO$_4$ hydrate crystals placed outside of the crucible. At high temperatures, MgSO$_4$ hydrate releases water, which facilitates the growth of larger crystals, but also introduces impurities into the crystals. The sealed quartz tube was then placed in a furnace and heated to 1020 $^{\circ}$C, followed by cooling at a rate of 6 $^{\circ}$C per hour for crystallization. The cooling rate was much faster than what had been used for the growth of Hg1201 (2 $^{\circ}$C per hour \cite{ZhaoAdv2006}), but it was chosen because a slower cooling rate was found to favor the formation of a mixture of single- and double-layer phases in the final product. The best crystals obtained in this way had a size of up to 2 mm along the CuO$_2$ planar dimensions and a thickness of about 0.1 mm. One such as-grown Hg1212 crystal is shown in Fig.~\ref{Fig0}(b). The crystals was characterized using both SQUID magnetometry (Fig.~\ref{Fig0}(a)) and X-ray Laue diffraction (Fig.~\ref{Fig0}(c)) to determine its quality. The magnetic susceptibility measurement showed a fairly sharp superconducting transition, with a 10-90\% transition width of 15 K centered at $T_\mathrm{c} \sim 90$ K. Combined with the sharp Laue pattern (Fig.~\ref{Fig0}(c)), the result indicates a reasonably good quality of the as-grown crystal.

A problem with this method is that the quartz tubes have a relatively high chance of exploding during the growth, because the required vapor pressure generated by the excess amount of HgO at high temperatures approaches the pressure limit that can be sustained even by the best quartz tubes. Moreover, we were unable to reach the required pressures for the synthesis of Hg1223 with this method. In order to overcome this problem, we therefore built a special high-pressure container to serve as a second layer of encapsulation, in order to prevent the sealed quartz tube from exploding.

\begin{figure}
\includegraphics[width=2.7in]{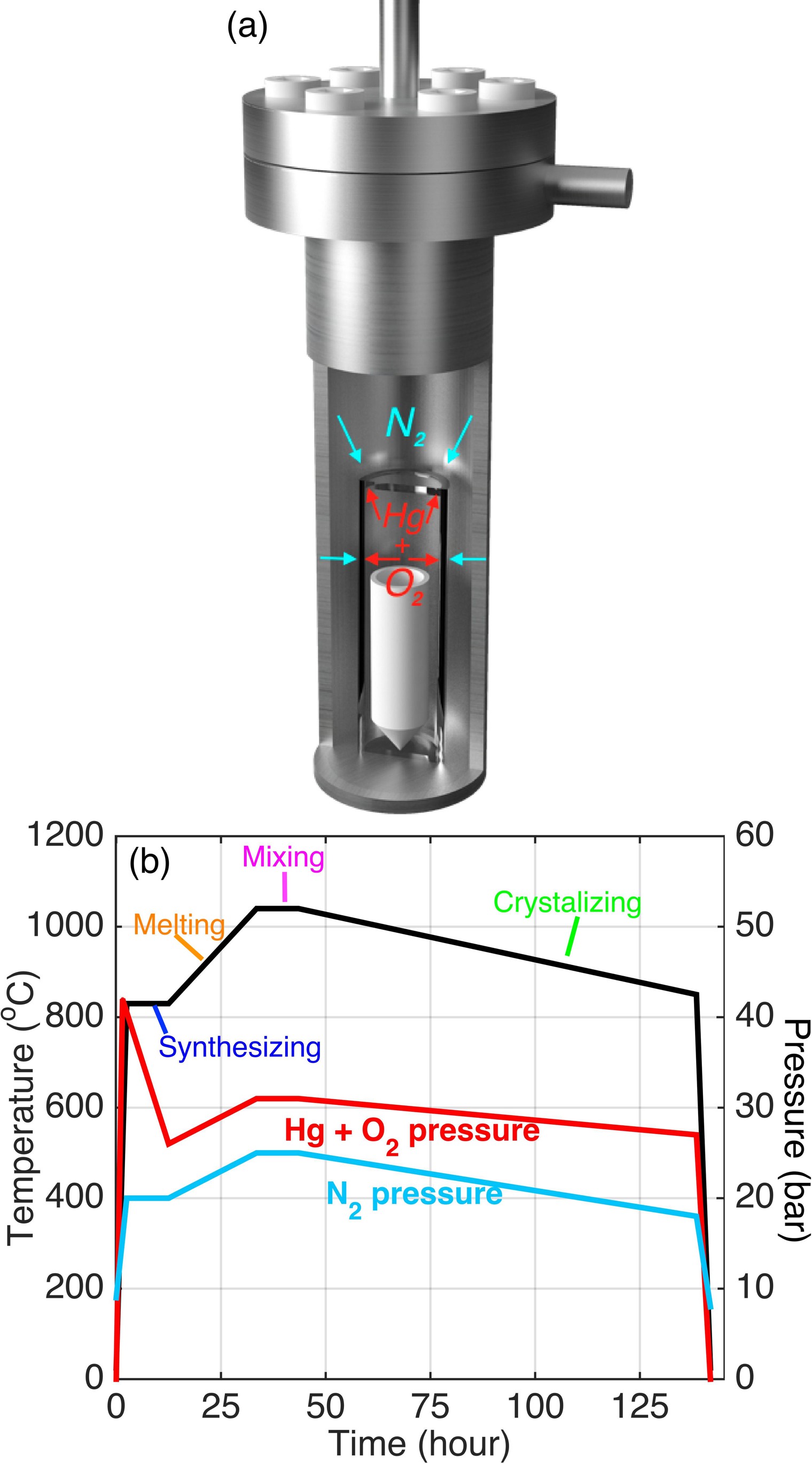}
\caption{\label{Fig1}
(a) Sketch of custom-designed furnace with two layers of encapsulation. As temperature increases, the heated nitrogen that surrounds the sealed quartz tube provides a counter pressure to prevent the quartz tube from exploding.
(b) Black: Temperature profile for the crystal growth of Hg1212 and Hg1223. Red: Total pressure of Hg and O$_2$ inside the quartz tube, generated by a mixture (see text) that contains 0.75 g of excess HgO, estimated from the ideal-gas equation. In the ``synthesis'' step, the pressure decreases because some of the HgO reacts with the rest of the materials. Blue: N$_2$ pressure read from the gauge.
}
\end{figure}

We now describe this new growth procedure in detail. The growth of Hg1212/Hg1223 crystals involved two steps: the initial preparation of precursor and the subsequent synthesis/crystal growth in a high-pressure furnace [Fig.~\ref{Fig1}(a)]. The precursor was prepared in the following manner, with no calcium included in order to avoid contamination due to the formation of carbonates. Powders of 99.999\% Ba(NO$_3$)$_2$ (Alfa Aesar) and 99.999\% CuO (Alfa Aesar) were firstly dried at 300 $^{\circ}$C to eliminate H$_2$O before weighing, and then mixed and ground in an agate mortar in stoichiometric amounts (different for Hg1212 and Hg1223). The mixture was then sintered in a horizontal tube furnace at 920 $^{\circ}$C for 10 hours, and then cooled to 200 $^{\circ}$C inside the furnace. The nitrogen in Ba(NO$_3$)$_2$ escaped in the form of gaseous NO$_2$, while the precursor Ba$_2$Cu$_2$O$_y$/Ba$_2$Cu$_3$O$_y$ remained. During the sintering procedure, the furnace was flooded with flowing high-purity oxygen in order to supply sufficient oxygen to the precursor and to prevent H$_2$O and CO$_2$ from entering the furnace, as these molecules are known to be the main sources of contamination \cite{AntiSupSciTec2002}. After the precursor was taken out from the furnace, it was immediately transferred into an argon glovebox, in which it was mixed with a stoichiometric amount of CaO and an excess amount of HgO. The mixture was put into a zirconia crucible, which was then sealed into an evacuated thick-walled (3 mm thickness) quartz tube. By carefully tuning the amount of excess HgO in the mixture, we were able to find the optimal range for the two different phases. For Hg1212, 2.04 g of Ba$_2$Cu$_2$O$_y$ precursor was mixed with stoichiometric amounts of CaO powder (0.25 g) and HgO powder (0.95 g), as well as an additional 0.7-0.75 g of HgO. For Hg1223, besides using its own stoichiometric amounts of precursor, CaO, and HgO, the optimal additional amount of HgO was 0.8-0.85 g.
The additional amounts of HgO were 2-3 times of that used for the growth of Hg1201 (0.3 g) \cite{ZhaoAdv2006}.  Altogether, for crystal growth of the Hg-family of HTSCs using the flux method, the required Hg pressure increases with the number of CuO$_{2}$ layers in a moderate fashion, a finding that differs from earlier conclusions based on the syntheses of polycrystalline samples\cite{XuePhC1997,KnizekPhC2002}.

The subsequent synthesis and crystal growth took place in a custom-built high-pressure furnace. The sealed quartz tube was surrounded by a protecting open tube of stainless steel (not shown here), and placed into a cylinder made of a nickel-based alloy, as depicted in Fig.~\ref{Fig1}(a). As the second layer of encapsulation, the nickel cylinder was designed with the capability of sustaining pressures up to 30 bar at 1100 $^{\circ}$C. It was filled at room temperature with an inert gas such as nitrogen and sealed with a gasket, in order to obtain a counter pressure to prevent the quartz tube from exploding. In previous syntheses and crystal growths of multi-layered Hg-family of compounds, the phases formed at temperatures that ranged from 800 $^{\circ}$C to 900 $^{\circ}$C (Ref. \onlinecite{PutilinPhC1993, MengPhysC1993, ColsonPhysC1994, XuePhysC1997, KnizekPhC2002}), and the melting point of the flux was in the range between 1000 $^{\circ}$C and 1070 $^{\circ}$C (Ref. \onlinecite{PelloquinPhysC1997,WisniewskiPRB2000}). Figure~\ref{Fig1}(b) shows the optimized temperature and pressure profiles covering a total of four stages. The nitrogen gas  pressure was read from a gauge attached to the nickel cylinder, whereas the Hg and O$_2$ pressure were rough estimations. In the synthesis stage, the furnace was rapidly heated to 830 $^{\circ}$C and then kept at this temperature for 10 hours. During this time, the pressure inside the quartz tube first rapidly increased as a result of the decomposition of HgO, and then decreased due to the formation of the desired phases as the precursor reacted with CaO, gaseous Hg and O$_2$. The pressure differential between the inside and outside of the quartz tube briefly reached maximum, which could result in explosion if the quartz tube and/or the seal was faulty. Contaminations in the precursor should be minimized, since they lead to increased explosions by hampering the reaction. In the following three stages, the temperature was first raised to 1040 $^{\circ}$C above the melting point of the flux, then kept constant for 10 hours, and finally slowly decreased at rate of 2 $^{\circ}$C/hr to achieve crystallization. Although the pressure differential was much smaller than the peak value, explosions could still occur since the strength of the quartz tube was diminished above 1000 $^{\circ}$C, and because faults could continually develop with time. In order to minimize explosions during these stages, the quartz tubes were annealed at 1250 $^{\circ}$C for 2 days before they were used for the growths; tubes of lesser quality usually exhibited visible cracks after this anneal and were discarded.

\begin{figure}
\includegraphics[width=3.375in]{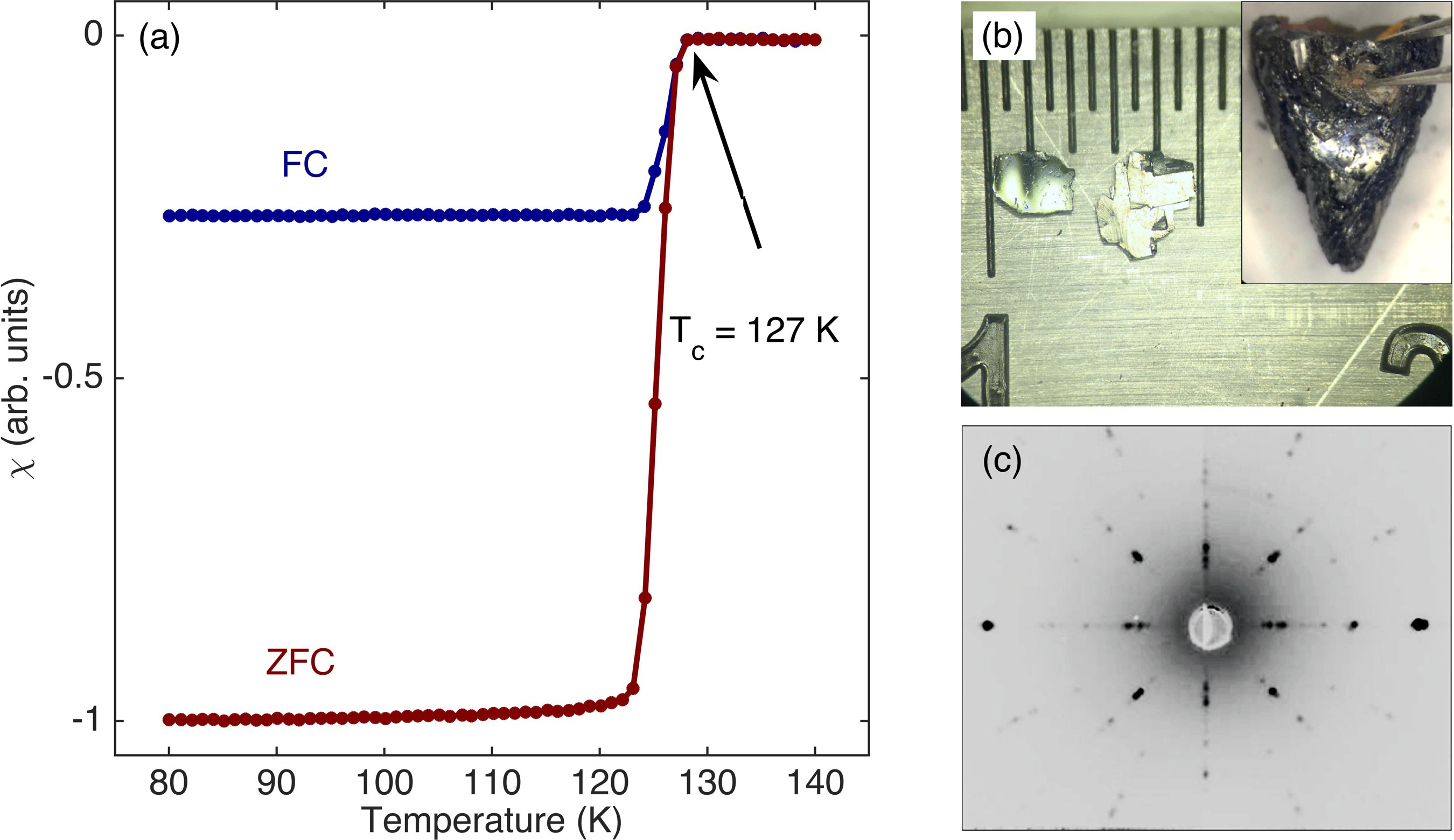}
\caption{\label{Fig2}
(a) Magnetic susceptibility measurements of an as-grown Hg1223 single crystal in a 5 Oe $c$-axis field, after zero-field cooling (red) and field cooling (blue). (b) Photograph of single crystals with naturally-cleaved surfaces. Inset: Raw product from the growth containing single crystals. (c) $c$-axis X-ray Laue pattern for a Hg1223 single crystal.
}
\end{figure}

\section{SAMPLE CHARATERIZATION}

\begin{figure}
\includegraphics[width=2.7in]{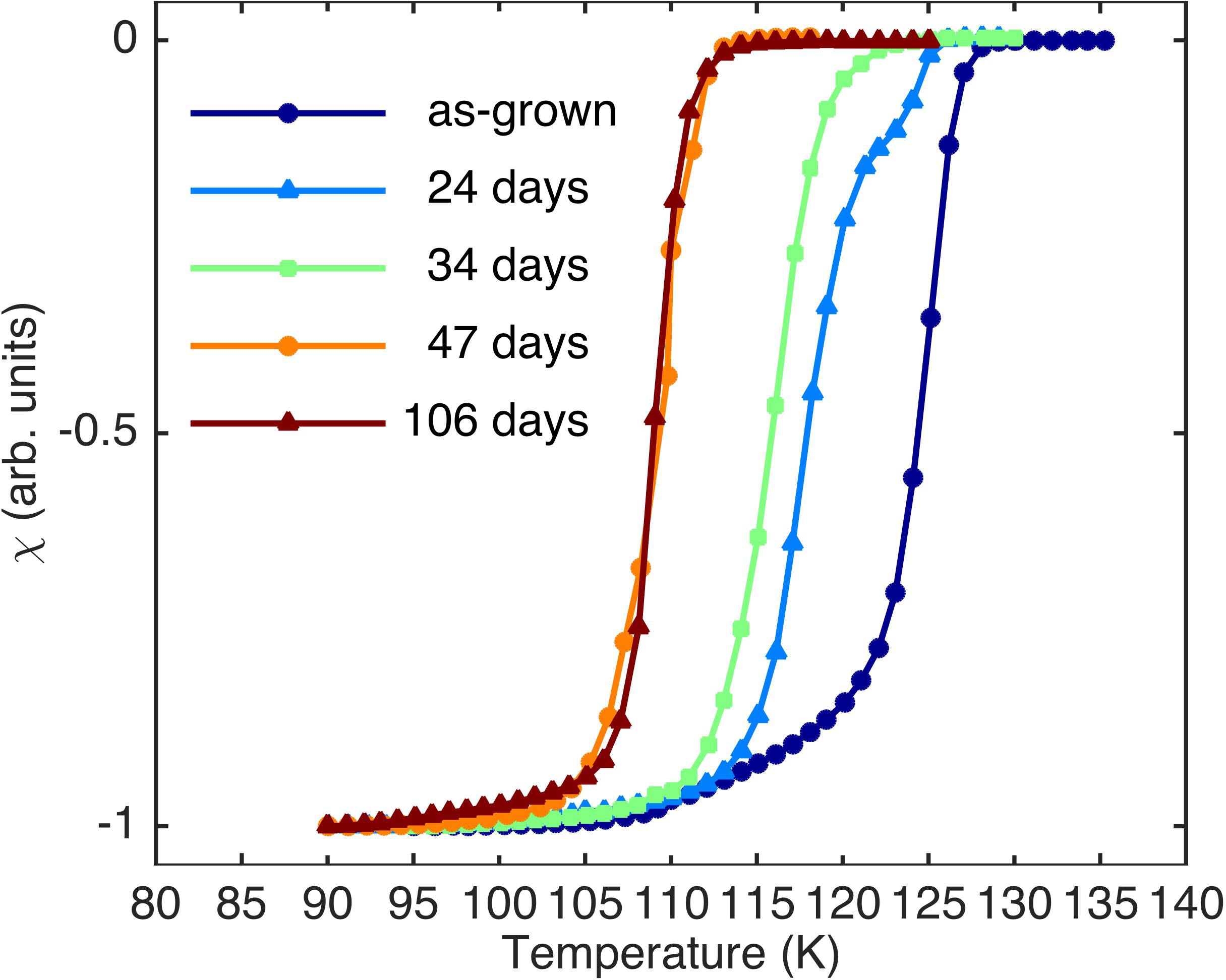}
\caption{\label{Fig3}
Anneal history of a Hg1223 crystal in air at 480 $^{\circ}$C. The measurements were performed in a 5 Oe $c$-axis field after zero-field cooling.
}
\end{figure}

Using the two-layer encapsulation method, we were able to obtain high-quality single crystals of Hg1212 and Hg1223. Successful synthesis yields a black final product with single crystals embedded inside, with typical $ab$-plane dimensions of several millimeters and a thickness of 0.3 mm, as is shown in Fig.~\ref{Fig2}(b). The high quality and homogeneity of some of the best as-grown crystals are demonstrated by sharp X-ray Laue patterns [Fig.~\ref{Fig2}(c)] and sharp superconducting transitions (of about 5 K in width), both in field-cooled and zero-field-cooled magnetic susceptibility measurements [Fig.~\ref{Fig2}(a)]. Some as-grown crystals exhibit high homogeneity, as shown in Fig.~\ref{Fig2}(a), whereas others show rather broad transition widths, as shown in Fig.~\ref{Fig3}. In the latter case, post-growth anneals proved useful to homogeneously tune the oxygen content, and here we take an Hg1223 crystal as example. Figure~\ref{Fig3} shows the annealing history of a Hg1223 crystal in air at 480 $^{\circ}$C, in which the transition was tuned from an initial value of $T_\mathrm{c} \sim 127$ K with broad transition, to $T_\mathrm{c} \sim 110$ K with sharp transition, after 47 days. No further changes were observed after an additional 60-day anneal. The superconducting transition temperature of Hg1212 is tuned to $T_\mathrm{c} \sim100$ K under the same annealing condition. However, we found for crystals with a large amount of growth defects that their $T_\mathrm{c}$ values (and hence their hole concentrations) could not be uniformly tuned by annealing, presumably because the defects preclude the movement of oxygen atoms. This observation allowed us to use post-growth anneals to select crystals of high quality.  Overall, doping control into the underdoped regime is achievable in our Hg1212 and Hg1223 single crystals. We established that the typical annealing time is about 50 days, which is longer than for underdoped Hg1201 \cite{BarisicPRB2008}. This difference can be attributed to the smaller in-plane lattice constants of Hg1212 and Hg1223 \cite{AntiSupSciTec2002}, which renders it more difficult for the interstitial oxygen atoms to enter or to escape from the Hg-O layers. Due to the long anneal times, we are still in the process of improving our knowledge on the doping control of Hg1212 and Hg1223, especially in the very underdoped and the overdoped regimes.

\begin{figure}
\includegraphics[width=3.375in]{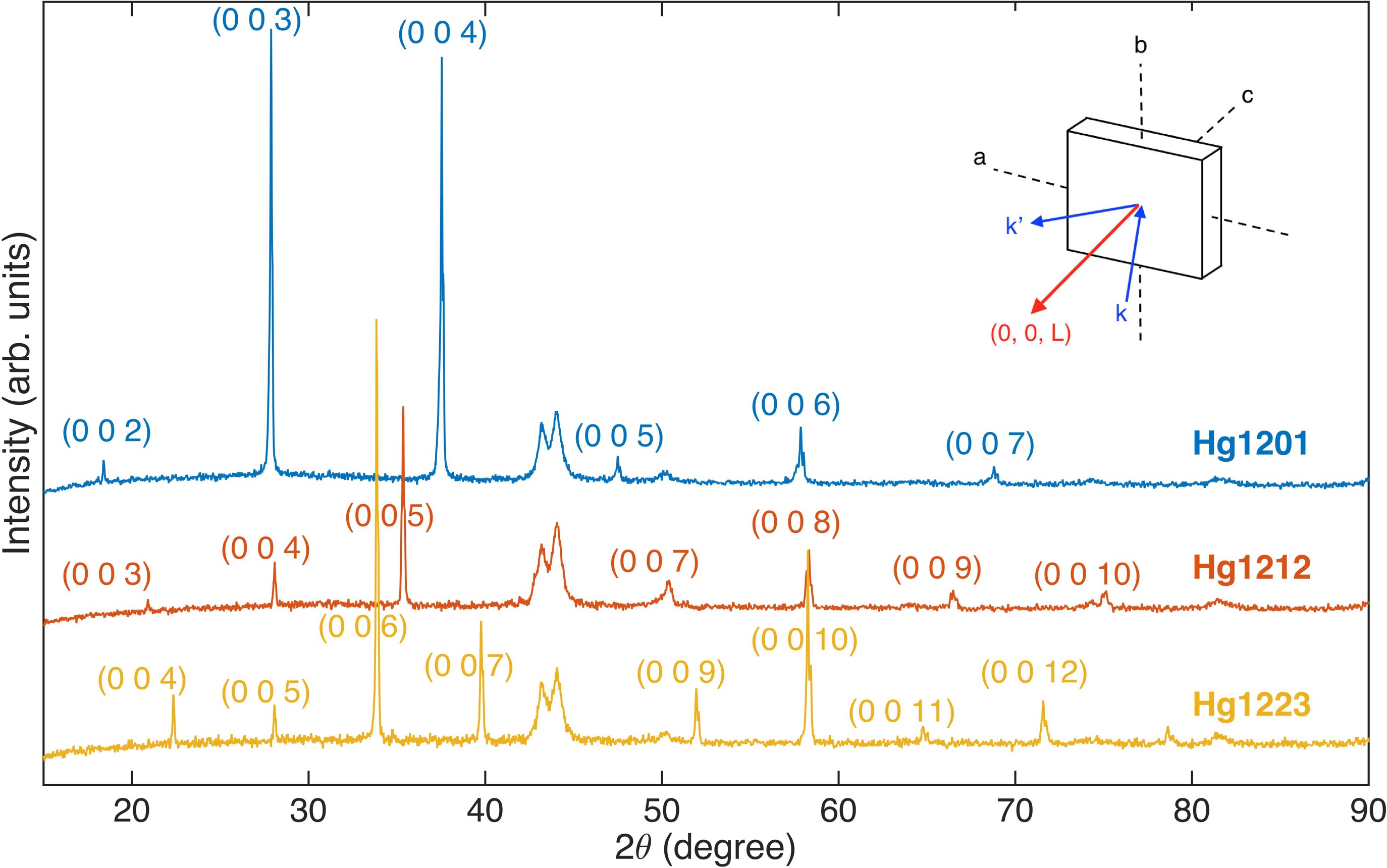}
\caption{\label{Fig4}
Single-crystal X-ray diffraction data for Hg1201, Hg1212 and Hg1223, with momentum transfer along the $c$ axis. The common peaks around 43 degrees originate from the sample plate. The (0 0 6) reflection of Hg1212 and the (0 0 8) reflection of Hg1223 are too weak and/or too close to these peaks to be visible. Inset: Illustration of the diffraction geometry.
}
\end{figure}

The crystal growth of the multi-layered Hg-family compounds inevitably suffers from inter-growth problems, similar to those of other cuprates \cite{KnizekPhC1993,HolePhC1992}. Crystals with different numbers of CuO$_{2}$ layers are not distinguishable by their appearance, and because as-grown samples tend to be underdoped to varying degrees, they cannot be distinguished by their $T_\mathrm{c}$ either. However, since they possess similar in-plane lattice constants ($a$ and $b$), but distinct out-of-plane lattice constants ($c$), single-crystal X-ray diffraction with momentum transfer parallel to the $c$ axis is able to distinguish them. Using the scattering geometry displayed in the inset of Fig.~\ref{Fig4}, characteristic diffraction patterns are obtained for Hg1201, Hg1212, and Hg1223 (Fig.~\ref{Fig4}), indicative of the increasing $c$ =  9.52 \r{A}, 12.70 \r{A} and 15.86 \r{A}, respectively. In this way, we are able to determine the type of single crystals in a non-destructive fashion. According to measurements of hundreds of crystals, the majority phase of a growth corresponds to the mass ratio in the starting materials. We did not find crystals with $c$ larger than that of Hg1223, which implies that even higher Hg and O$_2$ partial pressures are required for the synthesis of compounds with $n>3$.

\begin{figure}
\includegraphics[width=3.375in]{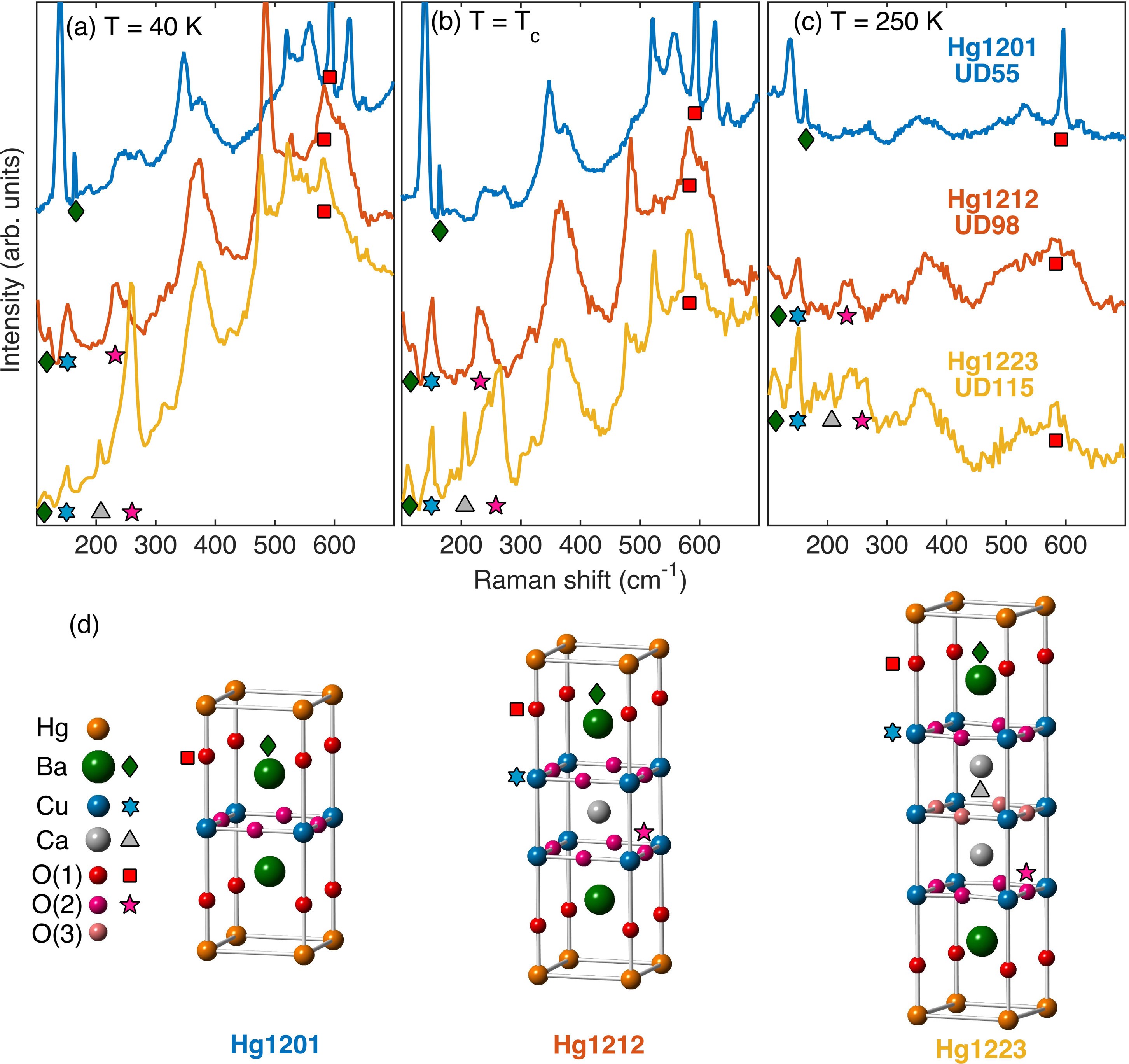}
\caption{\label{Fig5}
(a-c) $A_{1g}$ Raman spectra for Hg1201, Hg1212 and Hg1223 measured at different temperatures. The data are normalized and offset for clarity. Symbols indicate different $A_{1g}$ phonon modes and are coded with the assigned dominant atomic motion (see text) in the crystal structures in (d).
}
\end{figure}

In addition X-ray diffraction, we used Raman spectroscopy to characterize our samples, by taking advantage of the fact that characteristic Raman-active phonon peaks are distinctly different for the members of the Hg-family cuprates. Figure~\ref{Fig5} displays Raman spectra obtained on three single crystals in the underdoped (UD) regime: Hg1201 ($T_\mathrm{c}$ = 55 K), Hg1212 ($T_\mathrm{c}$ = 98 K), and Hg1223 ($T_\mathrm{c}$ = 115 K). The data were collected in a confocal backscattering geometry using a Horiba Jobin Yvon LabRAM HR Evolution spectrometer, equipped with 600 gr/mm gratings and a liquid-nitrogen-cooled CCD detector. The 633 nm line from a He-Ne laser was used for excitation. The linear polarization of both the incident and the scattered photons was restricted to be parallel to [1 1 0]  (a diagonal of the $ab$ plane), so that optical phonons that correspond to the $A_{1g}$ representation of the $D_{4h}$ group were detected. Because phonon peaks generally become sharper at lower temperature, the measurements were performed at low temperatures with the samples installed inside an ultrahigh-vacuum liquid-helium flow cryostat.

According to structural analysis, Hg1201, Hg1212, and Hg1223 ought to have a total of 2, 4, and 5 $A_{1g}$ optical phonons, respectively. However, all spectra displayed in Fig.~\ref{Fig5} feature more peaks than expected. The additional peaks are generally considered ``defect'' peaks that are indicative of a lower local symmetry than the nominal $D_{4h}$ point group with perfect lattice translational symmetry, and they will be discussed later. In order to first identify the regular $A_{1g}$ phonons, we attempt to assign each of them to the motion of primarily one type of atom on a given site. This is only an approximation, because the phonons' true eigenvectors are in principle superpositions of atomic motions that belong to the same irreducible representation of the symmetry group, and there might not be a dominant atomic displacement for each eigenvector. Nevertheless, we find that this approximation is reasonably good, because the phonon frequencies observed in the three compounds and summarized in Table~\ref{table:one} are quite similar, which suggests that the underlying eigenvectors are similar. Considering the fact that the three compounds have different atoms (and sites) that can participate in $A_{1g}$ phonons, this similarity can be naturally understood if each phonon primarily involves only one type of atomic motion.

The assignments in Table~\ref{table:one} follow the idea that the heavier atoms are expected to give rise to lower-energy modes. The individual modes are depicted by symbols placed next to the Raman spectra in Fig.~\ref{Fig5}(a-c), and next to the atoms in Fig.~\ref{Fig5}(d). For these $A_{1g}$ modes, the atomic displacements are always along the $c$ axis, and it is understood that the same type of atom in the upper and the lower halves of the displayed unit-cell structure must move in opposite ways, such that the eigenvector is invariant under all symmetry operations, including space inversion. We provide the following additional reasoning behind our assignment: (1) For Hg1201, we assign the two sharpest peaks at 163 and 594 cm$^{-1}$ to displacements of the heavier Ba and the lighter apical oxygen atom O(1), respectively. All other peaks are considerably broader and their intensities less polarization-dependent, so they are most likely ``defect'' peaks. (2) For Hg1212 (and Hg1223), the corresponding Ba and O(1) modes likely correspond to the peaks at about 119(112) and 583 cm$^{-1}$, whereas the additional Cu and planar oxygen O(2) modes probably correspond to the peaks at 152 and 232(264) cm$^{-1}$, according to an earlier report based on polycrystalline samples \cite{ZhouPhC1996}. Some of the nearby features, such as the shoulder at about 245 cm$^{-1}$ next to the O(2) peak in Hg1223, are ``defect'' peaks because our additional tests (not shown) indicate that they are less sensitive to changes in photon polarization than the regular phonons. (3) Finally, only Hg1223 has a fifth $A_{1g}$ phonon related to the motion of Ca, and it is considered to be the peak at 206 cm$^{-1}$ because this peak is only present in the spectra of Hg1223. In fact, this peak is already quite visible in the 250 K spectrum and even at room temperature, so that it can be utilized to distinguish between Hg1212 and Hg1223 without the need to cool a sample. It is relatively straightforward to distinguish both Hg1212 and Hg1223 from Hg1201: the former have more peaks in the 200-300 cm$^{-1}$ energy range and a less pronounced O(1) phonon peak near 590 cm$^{-1}$.

\begin{table}[tbp]%\footnotesize Involved atoms of vibrating mode (Energy in cm$^{-1}$)
% \centering
%\begin{tabular}{|c|c|c|}\hline\hline
\begin{tabular}{c|c|c|c|c|c}\hline\hline
Compound & \multicolumn{5}{c} {Involved atoms of phonons (Energy in cm$^{-1}$)} \\
\hline
Hg1201 &Ba  163  &  & &  &O(1) 594 \\
\hline
Hg1212  &Ba  119 &Cu  152 & &O(2)  232 &O(1) 583 \\
\hline
Hg1223 &Ba  112 &Cu  152 & Ca 206 &O(2) 264 &O(1) 583\\
\hline\hline
\end{tabular}
\caption{\label{table:one}
The assignment of $A_{1g}$ phonons for Hg1201, Hg1212 and Hg1223. The energies of the phonons are determined from the spectra measured at the respective $T_\mathrm{c}$ of the three crystals: 55, 98, and 115 K.
}
\end{table}

Last, but not least, the Raman spectra measured at 250 K already contain additional structures that cannot be attributed to the regular $A_{1g}$ phonons. While they might be due to defect peaks or multi-phonon scattering, it is surprising to see that many of them become much more pronounced upon cooling, and that they share approximately the same energies in all three compounds. Some of the peaks are even only discernible at low temperatures, such as the one at about 520 cm$^{-1}$. Since no structural phase transitions have been reported for these compounds, these temperature-dependent anomalous peaks might be related to the formation of short-range order that causes the lattice to locally deform. Indeed, additional Raman-active modes can be expected if the local point-group symmetry is lowered, and/or if the lattice translational symmetry is broken.  These possibilities correspond well with the discovery of charge order in the high-$T_\mathrm{c}$ cuprates \cite{KeimerNature2015}, including Hg1201 \cite{TabisNatCom2014,TabisPRB2017}, and with torque magnetometry results that suggest a lowering of the global point-group symmetry in Hg1201 within the pseudogap phase \cite{Murayama2018}. Also, the cuprates are known to be intrinsically inhomogeneous due to in-built structural stress \cite{Krumhansl1992}, with the possibility of distinct local environments on the nanoscale, as suggested, e.g., by crystal-field-excitation spectroscopy \cite{Mesot1993} and refinement of the atomic pair density function obtained from neutron-powder-diffraction data \cite{Billinge1993}. Nuclear quadrupole resonance also shows a wide distribution of local electric field gradients in most cuprates, including Hg1201 \cite{Rybicki2009}. Interestingly, the observed additional peaks are quite similar to the phonon back-folding behavior recently reported for an insulating low-dimensional spin system \cite{WangPRB2017}. Whether such local lattice deformations, namely the spontaneous development of intrinsic structural inhomogeneity upon cooling, is generic to the high-$T_\mathrm{c}$ cuprates (or to low-dimensional spin-1/2 systems) deserves further investigation.

To summarize, we have successfully synthesized sizable single crystals of Hg1212 and Hg1223. We have demonstrated by a variety of methods that our samples are of high quality, and the doping level can be controlled to some extent by post-growth anneals. While inter-growth problems appear to be inevitable, we are able to distinguish the different crystals by using single-crystal X-ray diffraction and Raman spectroscopy. Raman spectroscopy has also allowed us to characterize the lattice dynamics. The data suggest the formation of short-range order at low temperature, along with local lattice deformations. The newly available crystals of the double- and triple-layer Hg-family HTSCs can be expected to enable new directions in the study of the cuprate HTSCs.

\begin{acknowledgments}
We wish to thank Weiliang Yao, Hongjie Chen and Yiqi Xie for their help with the synthesis and characterization, and D. Pelc for comments on the manuscript. The work at Peking University was supported by the National Natural Science Foundation of China (grant no. 11522429) and Ministry of Science and Technology of China (grant nos. 2018YFA0305602 and 2015CB921302). The work at the University of Minnesota was funded by the Department of Energy through the University of Minnesota Center for Quantum Materials under DE-SC-0016371.
\end{acknowledgments}

\preprint{Preprint}

\bibliography{reference}% Produces the bibliography via BibTeX.

\end{document}